\documentclass[12pt]{iopart}
\expandafter\let\csname equation*\endcsname\relax
\expandafter\let\csname endequation*\endcsname\relax

\usepackage{amsmath,amsfonts,amssymb,amsthm}
\usepackage{graphicx}
\usepackage{color}
\usepackage{hyperref}

\def\softd{{\leavevmode\setbox1=\hbox{d}%
\hbox to 1.05\wd1{d\kern-0.4ex{\char039}\hss}}}
\def\softt{{\leavevmode\setbox1=\hbox{t}%
\hbox to \wd1{t\kern-0.6ex{\char039}\hss}}}

\newtheorem{theorem}{Theorem}[section]

\newtheorem{remark}{Remark}[section]

\definecolor{Myred}{cmyk}{0.0,1.0,1.0,0.00}
\renewcommand{\theequation}{\arabic{section}.\arabic{equation}}

\begin{document}
\title[Magnetic Schr\"{o}dinger operators with radially symmetric magnetic field]
{Magnetic Schr\"{o}dinger operators with radially symmetric magnetic field and radially symmetric electric potential}

\author{Diana Barseghyan}
\address{Department of Mathematics, University of Ostrava, 30. dubna 22, 70103 Ostrava, Czech Republic}
\address{Nuclear Physics Institute, Academy of Sciences of the Czech Republic,
Hlavn\'{i} 130, 25068 \v{R}e\v{z} near Prague, Czech Republic}
\ead{dianabar@ujf.cas.cz, diana.barseghyan@osu.cz}

\author{Fran\c{c}oise Truc}
\address{ Institut Fourier, UMR 5582 du CNRS Universite de Grenoble I, BP 74, 38402 Saint-Martin d'Heres, France, Bureau 111 }
\ead{francoise.truc@ujf-grenoble.fr}

\begin{abstract}
The aim of the paper is to derive spectral estimates on the eigenvalue moments of the magnetic Schr\"{o}dinger operators defined on the two-dimensional disk with a radially symmetric magnetic field and radially symmetric electric potential.

\vspace{2pc} \noindent{\it Keywords}:  Eigenvalue bounds, radial
magnetic field, Lieb-Thirring inequalities, discrete spectrum.
\end{abstract}

\maketitle

\section{Introduction} \label{s: intro}
\setcounter{equation}{0}
Let us consider a particle in a bounded domain $\Omega$ in $\mathbb{R}^2$ in the presence of a magnetic field $B$ and an electric potential $V$.  We define the 2-dimensional magnetic Schr\"{o}dinger operator associated 
to this particle as follows:

Let $A$ be a magnetic potential associated to $B$, i.e. a smooth real valued-function on $\Omega \subset \mathbb{R}^2$ verifying   $\mathrm{rot}\, (A)=B$ and $V\ge0$ be a bounded measurable potential defined on $L^2(\Omega)$. 
The magnetic Schr\"{o}dinger operator is initially defined  on $ C_0^\infty (\Omega)$ by
$H_\Omega (A, V) = (i\nabla + A)^2-V$.

The case when the magnetic field is not constant can be motivated by anisotropic superconductors (see for instance \cite{CDG95}) or the liquid crystal theory.

Assuming some regularity conditions (RC) on $A$, namely : the magnetic field $B$ is $\in L_{\mathrm{loc}}^\infty(\Omega)$ and the corresponding magnetic potential $A$ is $\in L^\infty (\Omega)$, we get that the magnetic Sobolev norm $\|(i \nabla+A) u\|_{L^2(\Omega)},\,\,u\in \mathcal{H}_0^1 (\Omega)$, is closed and equivalent to the non-magnetic one which means that they both have purely discrete spectrum. Thus using the boundedness of the potential $V$  the self-adjoint Friedrichs extension of $H_\Omega (A, V)$ initially defined on $ C_0^\infty (\Omega)$ 
 has a purely discrete spectrum.

In the paper we also consider the case when the magnetic field grows to infinity as the variable approaches the boundary 
and has a non zero infimum \begin{equation}\label{grow}B(z)\to\infty\quad\text{as}\quad z\to \partial\Omega \quad\text{and}\quad
K:=\inf\,B(z)>0.\end{equation} In view of the lower bound $$(H_\Omega (A, V) (u), u)_{L^2(\Omega)}\ge \int_\Omega \left(B(z)-\|V\|_{L^\infty(\Omega)}\right) |u|^2(z)\,\mathrm{d}z,$$ one  again can construct the Friedrichs extension of $H_\Omega (A, V)$ initially defined on $C_0^\infty (\Omega)$. Moreover, it still has a purely discrete spectrum--\cite{T12}.

For simplicity, we will use for the Friedrichs extension the same symbol  $H_\Omega(A, V)$, and 
we shall denote the increasingly ordered sequence of  its eigenvalues by $\lambda_k =\lambda_k(\Omega, A, V)$.

The  purpose of  this paper is to establish bounds of the eigenvalue moments of such operators. 
Let us recall the following bound which was proved by Berezin, Li and Yau   for non-magnetic Dirichlet Laplacians on a domain  $\Omega$ in $\mathbb{R}^d$
-- \cite{Be72a, Be72b, LY83},
\begin{equation}
\label{Berezin bound}\sum_k(\Lambda-\lambda_k(\Omega,0, 0))_+^\sigma\le L_{\sigma,d}^{\mathrm{cl}}\,|\Omega|\,
\Lambda^{\sigma+\frac{d}{2}} \quad\text{for any}\;\;\sigma\ge1 \;\;\text{and}\;\;\Lambda>0\,,
\end{equation}
where $|\Omega|$ is the volume of $\Omega$, and the constant on the right-hand side,
\begin{equation}\label{semiclassical constant}
L_{\sigma,d}^{\mathrm{cl}}=\frac{\Gamma(\sigma+1)}{(4\pi)^{\frac{d}{2}}\Gamma
(\sigma+1+d/2)}\,,
\end{equation}
is optimal. Moreover,  for $0\le\sigma<1$, the bound (\ref{Berezin bound})  still exists, but with another constant on the right-hand side
-- \cite{La97} 
\begin{equation}\label{Laptev ineq.}
\sum_k(\Lambda-\lambda_k(\Omega,0, 0))_+^\sigma\le
2\left(\frac{\sigma}{\sigma+1}\right)^\sigma L_{\sigma,d}^{\mathrm{cl}}\,|\Omega|\,
\Lambda^{\sigma+\frac{d}{2}}\,,\quad 0\le\sigma<1\,.
\end{equation}
For Schr\"{o}dinger operators $H_\Omega (0, V)$ with the Dirichlet boundary conditions the following bound was proved by Lieb-Thirring --\cite{LT76} \begin{equation}\label{Lieb-Thirring ineq.}
\sum_{\lambda_k(\Omega, 0,V)\le0}|\lambda_k(\Omega, 0, V)|^\sigma\le
L_{\sigma,d}^{\mathrm{cl}}\,\int_\Omega V^{\sigma+d/2}(x)\,\mathrm{d}x\,,\quad \sigma\ge3/2\,.
\end{equation}

The similar estimate for the Schr\"{o}dinger operators $H_\Omega (A, V)$ with the Dirichlet boundary conditions and with non-zero magnetic field takes place --\cite{LW00}
\begin{equation}\label{Laptev Weidl ineq.}
\sum_{\lambda_k(\Omega, A,V)\le0}|\lambda_k(\Omega, A, V)|^\sigma\le
L_{\sigma,d}^{\mathrm{cl}}\,\int_\Omega V^{\sigma+d/2}(x)\,\mathrm{d}x\,,\quad \sigma\ge3/2\,.
\end{equation}

In the magnetic case, due to the pointwise diamagnetic inequality which means that under  rather general assumptions on the magnetic potentials \cite{LL01}
$$
|\nabla|u(x)||\le|(i\nabla+A)u(x)|\quad\text{for a.a.}\;\; x\in\Omega\,,
$$
we get  that $\lambda_1(\Omega, A, 0)\ge\lambda_1(\Omega,0, 0)$. However, the estimate $\lambda_j(\Omega, A, 0)\ge\lambda_j(\Omega,0, 0)$ fails in general 
if $j\ge2$. Let us mention that,  nevertheless, momentum estimates are still valid for some values of the parameters. In particular, it was shown \cite{LW00} that 
the sharp bound (\ref{Berezin bound}) holds true for arbitrary magnetic fields provided $\sigma\ge\frac{3}{2}$, and  for constant magnetic fields if 
$\sigma\ge1$-- \cite{ELV00}, \cite{KW15}. In the two-dimensional case  the bound (\ref{Laptev ineq.}) holds true for constant magnetic fields if $0\le\sigma<1$, and the constant on 
the  right-hand side cannot be improved --\cite{FLW09}.

In the present work we study the magnetic Schr\"{o}dinger operators $H_\Omega(A, V)$ defined on the two- dimensional disk $\Omega$ 
centered in zero and with radius $r_0>0$, with a radially symmetric magnetic field $B(x)=B(|x|)$ and electric potential $V=V(|x|)\ge0$.  Our  aim is to extend 
a sufficiently precise Lieb-Thirring type inequality to this situation.  A similar problem was studied recently for magnetic Dirichlet Laplacians
in \cite{BEKW16}, but under very strong restrictions on the growth of the magnetic field. 

Let us also mention that some estimates on the counting function of the eigenvalues of the magnetic Dirichlet Laplacian on a disk
 were established in \cite{T12}, in the case where the field is  radial and satisfies some growth condition near the boundary.

\bigskip \section{Main Result}\setcounter{equation}{0}
Inspired by the weighted one- dimensional Lieb-Thirring type inequalities \cite{EF08} we establish the weighted eigenvalue bound for the operator $H_\Omega(A, V)$ in terms of the magnetic and electric potentias $B$ and $V$. The following theorem holds true:  \begin{theorem}Let $H_\Omega(A, V)$ be the magnetic Schr\"{o}dinger operator with the Dirichlet boundary conditions defined on the disk $\Omega$ of radius equal to $r_0$ centered at the origin with a radial magnetic field $B(x)=B(|x|)$ and electric potential $V=V(|x|)\ge0$. 
Let us assume the validity of the conditions (RC) or the validity of (\ref{grow}). 
Then for any $0<\varepsilon\le3/4,\, 0\le\alpha<1$ and $\sigma\ge(1-\alpha)/2$, the following inequality holds 
\begin{multline}\mathrm{tr}\left(H_\Omega(A, V)\right)_-^\sigma\le\frac{2 r_0 L_{\sigma+1/2, \alpha}}{\sqrt{1-\varepsilon}}\int_0^{r_0}\left(\left(\frac{1}{\varepsilon}-1\right)\frac{1}{r^2}\left(\int_0^r s B(s)\,\mathrm{d}s\right)^2+V(r)-\frac{1}{4r^2} \right)_+^{\sigma+1+\alpha/2}\,r^\alpha\,\mathrm{d}r\\+\frac{L_{\sigma, \alpha}}{\sqrt{1-\varepsilon}} \int_0^{r_0}\left(\left(\frac{1}{\varepsilon}-1\right)\frac{1}{r^2}\left(\int_0^r s B(s)\,\mathrm{d}s\right)^2+V(r)-\frac{1}{4 r^2} \right)_+^{\sigma+(1+\alpha)/2}\,r^\alpha\,\mathrm{d}r\\\label{LTineq.}+L_{\sigma, \alpha} \int_0^{r_0}\left(V(r)-\frac{1}{r^2}\left(\int_0^r s B(s)\,\mathrm{d}s\right)^2\right)_+^{\sigma+(1+\alpha)/2}\,r^\alpha\,\mathrm{d}r\,,\end{multline}
where $L_{\sigma+1/2, \alpha}$ and $L_{\sigma, \alpha}$ are some constants.
\end{theorem}
\begin{remark}
If  $0\le\sigma<3/2$ then even for magnetic Laplacians (\ref{Berezin bound}) type inequality is known only for constant magnetic fields.
\end{remark}

\begin{remark}
If  $\mathrm{sup}_{r<r_0}\left(V(r)-\frac{1}{4r^2}\right)<-A^2/3$, where $A$ is:
$\mathrm{sup}_{r<r_0}\frac{1}{r}\int_0^r s B(s)\mathrm{d}s<\infty$ then we can choose $\varepsilon\ge3/4$ such that the first two terms of the right hand side of (\ref{LTineq.}) be  equal to zero.
So we decrease the order of the potential $V$ in Lieb-Thirring bound (\ref{Laptev Weidl ineq.}) from $\sigma+1$ to $\sigma+(1+\alpha)/2<\sigma+1$. 
\end{remark}

\begin{proof} We begin by recalling the standard partial wave decomposition :--\cite{E96} 
$$L^2(\Omega,  dx )=\bigoplus_{m=-\infty}^\infty L^2((0, r_0), 2\pi r dr)$$ 
 $$f \rightarrow (...,f_1,f_0,f_1,....) \quad \rm{with}\quad f(r,\theta ) = \sum_{m=-\infty}^\infty e^{im\theta} f_m (r)\ .$$ 
Choosing the radial gauge $ A(r,\theta) = (-a(r) sin\theta, a(r) cos\theta) $ where 

 $$  a(r) := \frac{1}{r}\int_0^r s B(s)\,\mathrm{d}s, $$ we get that the operator $H_\Omega (A, V)$ acts on $\bigoplus_{m=-\infty}^\infty L^2((0, r_0)$
 as follows

 $$H_\Omega(A, V)=\bigoplus_{m=-\infty}^\infty h_m(B, V),$$ 
where the operators $h_m(B, V)$ are the Friedrichs extension of the closures of the quadratic forms
$$Q(h_m(B, V))[u]= 2\pi \int_0^{r_0}  \left(\left|\frac{\mathrm{d}u}{\mathrm{d}r}\right|^2+\left(\frac{m}{r}-a(r)\,\mathrm{d}s\right)^2 |u|^2-V |u|^2\right)\,r\,\mathrm{d}r,$$ 
defined originally on $C_0^\infty(0, r_0)$, and acting on their domain as
$$h_m(B)=-\frac{\mathrm{d}^2}{\mathrm{d}r^2}-\frac{1}{r}\frac{\mathrm{d}}{\mathrm{d} r}+\left(\frac{m}{r}-a(r)\right)^2-V(r).$$ 
Employing the mapping $U: C_0^\infty(0, r_0)\to C_0^\infty(0, r_0)$ defined by $$(U f)(r)=\frac{1}{\sqrt{2\pi r}} f(r)$$ 
one gets the unitarily equivalence between the operators $h_m(B)$ and 
$l_m(B)=-\frac{\mathrm{d}^2}{\mathrm{d}r^2}-\frac{1}{4r^2}+\left(\frac{m}{r}-a(r)\right)^2-V(r)$ defined already on $L^2(0, r_0)$.
Thus we are going to consider the self-adjoint operators associated to the closures of the quadratic forms 
$$Q(l_m(B, V))[v]=\int_0^{r_0}  \left(\left|\frac{\mathrm{d}v}{\mathrm{d}r}\right|^2-\frac{1}{4r^2}|v|^2+\left(\frac{m}{r}-a(r)\right)^2 |v|^2-V(r)|v|^2\right)\,\mathrm{d}r,$$
defined originally on $C_0^\infty(0, r_0)$.

We have,  for any $0<\varepsilon<1$ and any $v\in C_0^\infty(0, r_0)$
\begin{multline*}Q(l_m(B, V))[v]=\int_0^{r_0}\left(\left|\frac{\mathrm{d}v}{\mathrm{d}r}\right|^2-\frac{1}{4r^2}|v|^2+\frac{m^2}{r^2}|v|^2-\frac{2m}{r}
a(r) |v|^2+a^2(r) |v|^2-V(r)|v|^2\right)\,
\mathrm{d}r\\\ge\int_0^{r_0}\left(\left|\frac{\mathrm{d}v}{\mathrm{d}r}\right|^2-\frac{1}{4r^2}|v|^2+\frac{m^2}{r^2}|v|^2-\frac{m^2 \varepsilon}{r^2}|v|^2-
\frac{1}{\varepsilon}a^2(r) |v|^2+a^2(r) |v|^2-V(r)|v|^2\right)\,
\mathrm{d}r.
\end{multline*}
It follows from the above inequality that if $m\ne0$ and $0<\varepsilon\le3/4$
\begin{equation}\label{g}l_m (B, V)\ge g_{B, V}+\frac{(1-\varepsilon)m^2-1/4}{r_0^2},\end{equation}
where the operator $g_{B, V}$ is associated with the closure of the form $$Q(g_{B, V})[v]=\int_0^{r_0}\left(\left|\frac{\mathrm{d}v}{\mathrm{d}r}\right|^2-\left(\frac{1}{\varepsilon}-1\right)a^2(r)|v|^2-V(r)|v|^2\right)\,\mathrm{d}r$$ initially defined on $C_0^\infty(0, r_0)$.  

Let $\{\mu_k(B, V)\}_{k=1}^\infty$ be the set of the negative eigenvalues of $g_{B, V}$.
Due to the  minimax principle the inequality (\ref{g}) implies \begin{eqnarray}\nonumber\mathrm{tr}\left(\bigoplus_{m=-\infty}^\infty h_m(B, V)\right)_-^\sigma\le\sum_{m=-\infty,\,m\ne0}^\infty\mathrm{tr}\left(g_{B, V}+\frac{(1-\varepsilon) m^2-1/4}{r_0^2}\right)_-^\sigma\\\nonumber+\mathrm{tr} \left(-\frac{\mathrm{d}^2}{\mathrm{d} r^2}-\frac{1}{4 r^2}+a^2(r)-V(r)\right)_-^\sigma\\\nonumber\le\sum_{m=-\infty,\,m\ne0}^\infty\sum_{\mu_k(B, V)+\left((1-\varepsilon) m^2-1/4\right)/r_0^2\le0}\left|\mu_k(B, V)+\frac{(1-\varepsilon) m^2-1/4}{r_0^2}\right|^\sigma\\\nonumber+\mathrm{tr} \left(-\frac{\mathrm{d}^2}{\mathrm{d} r^2}-\frac{1}{4 r^2}+a^2(r)-V(r)\right)_-^\sigma\\\nonumber\le\sum_{k=1}^\infty\,\sum_{0<|m|\le\frac{\sqrt{|\mu_k(B, V)| r_0^2+1/4}}{\sqrt{1-\varepsilon}}}\left|\mu_k(B, V)+\frac{(1-\varepsilon) m^2-1/4}{r_0^2}\right|^\sigma\\\nonumber+\mathrm{tr} \left(-\frac{\mathrm{d}^2}{\mathrm{d} r^2}-\frac{1}{4 r^2}+a^2(r)-V(r)\right)_-^\sigma\\\nonumber\le\sum_{k=1}^\infty\left(\frac{2\sqrt{|\mu_k(B, V)| r_0^2+1/4}}{\sqrt{1-\varepsilon}}|\mu_k(B, V)|^\sigma\right)\\\nonumber+\mathrm{tr} \left(-\frac{\mathrm{d}^2}{\mathrm{d} r^2}-\frac{1}{4 r^2}+a^2(r)-V(r)\right)_-^\sigma\\\nonumber\le\frac{2 r_0}{\sqrt{1-\varepsilon}}{\sum_{k=1}^\infty}|\mu_k (B, V)|^{\sigma+1/2}+\frac{1}{\sqrt{1-\varepsilon}}\sum_{k=1}^\infty |\mu_k(B, V)|^\sigma\\\label{mu}+\mathrm{tr} \left(-\frac{\mathrm{d}^2}{\mathrm{d} r^2}-\frac{1}{4 r^2}+a^2(r)-V(r)\right)_-^\sigma.\end{eqnarray}
Let us extend the potential $-\left(\frac{1}{\varepsilon}-1\right)a^2(r)-V(r)$ to $\mathbb{R}_+$
by zero and denote the corresponding one dimensional Schr\"{o}dinger operator by $g^*(B, V)$.
Since $C_0^\infty(0, r_0)\subset C_0^\infty(\mathbb{R}_+)$ then by minimax principle for any $\delta>0$ \begin{equation}\label{nu}\sum_k|\mu_k(B, V)|^\delta\le\sum_k|\nu_k(B, V)|^\delta,\end{equation} where $\{\nu_k(B, V)\}_{k=1}^\infty$ are the negative eigenvalues of $g^*(B, V)$. 

Applying the Lieb-Thirring inequality --\cite{EF08} for any $\alpha\in[0, 1)$ and $\sigma\ge(1-\alpha)/2$ we get 
\begin{eqnarray}\nonumber\sum_{k=1}^\infty |\nu_k(B, V)|^{\sigma+1/2}\le L_{\sigma+1/2, \alpha} \int_0^{r_0}\left(\left(\frac{1}{\varepsilon}-1\right)a^2(r)+ V(r)-\frac{1}{4r^2}\right)_+^{\sigma+1+\alpha/2}\,r^\alpha\,\mathrm{d}r\,,\\\label{final}\sum_{k=1}^\infty |\nu_k(B, V)|^\sigma\le L_{\sigma, \alpha} \int_0^{r_0}\left(\left(\frac{1}{\varepsilon}-1\right)a^2(r)+ V(r)-\frac{1}{4r^2}\right)_+^{\sigma+(1+\alpha)/2}\,r^\alpha\,\mathrm{d}r\,,\\\nonumber\mathrm{tr}\left(-\frac{\mathrm{d}^2}{\mathrm{d} r^2}-\frac{1}{4 r^2}+a^2(r)-V(r)\right)_-^\sigma\le L_{\sigma, \alpha} \int_0^{r_0}\left(V(r)-a^2(r)\right)_+^{\sigma+(1+\alpha)/2}\,r^\alpha\,\mathrm{d}r.\end{eqnarray}
where $L_{\sigma+1/2, \alpha}$ and  $L_{\sigma, \alpha}$ are some constants.

This together with the estimates (\ref{mu})-(\ref{nu}) means 
\begin{eqnarray}\nonumber\mathrm{tr}\left(\bigoplus_{m=-\infty}^\infty h_m(B, V)\right)_-^\sigma\\\nonumber \le\frac{2 r_0 L_{\sigma+1/2, \alpha}}{\sqrt{1-\varepsilon}}\int_0^{r_0}\left(\left(\frac{1}{\varepsilon}-1\right)a^2(r)+V(r)-\frac{1}{4r^2} \right)_+^{\sigma+1+\alpha/2}\,r^\alpha\,\mathrm{d}r\\\nonumber+\frac{L_{\sigma, \alpha}}{\sqrt{1-\varepsilon}} \int_0^{r_0}\left(\left(\frac{1}{\varepsilon}-1\right)a^2(r)+V(r)-\frac{1}{4 r^2} \right)_+^{\sigma+(1+\alpha)/2}\,r^\alpha\,\mathrm{d}r\\\label{mne0}+L_{\sigma, \alpha} \int_0^{r_0}\left(V(r)-a^2(r)\right)_+^{\sigma+(1+\alpha)/2}\,r^\alpha\,\mathrm{d}r\,,\end{eqnarray}  which proves the theorem.

\end{proof}

\subsection*{Acknowledgements}
The work of D.B. is supported by Czech Science Foundation
(GA\v{C}R) within the project 17-01706S. D.B. also acknowledges a
support from the projects 01211/2016/RRC and the Czech-Austrian
Grant CZ 02/2017. F.T. is a member of the team of the ANR GeRaSic (G\'eom\'etrie spectrale, Graphes, Semiclassique).

D.B. appreciates the hospitality in Institut Fourier where the preliminary version of the paper was prepared.

\end{document}